\documentclass[prl,twocolumn,showpacs]{revtex4-1}
\usepackage{graphicx}
\usepackage{amsmath}

\renewcommand{\vec}[1]{\mathbf{#1}}
\newcommand{\Scal}{\mathcal{S}}

\begin{document}
\title{Fast thermalization and Helmholtz oscillations of 
  an ultracold Bose gas}
\author{D.J. Papoular$^{1}$, L.P. Pitaevskii$^{1,2}$, S. Stringari$^{1}$}
\affiliation{${}^{1}$INO-CNR BEC Center and Dipartimento di Fisica, 
  Universit\`a di Trento, 38123 Povo, Italy}
\affiliation{${}^{2}$Kapitza Institute for Physical Problems, Kosygina
2, 119334 Moscow, Russia}
\date{\today}

\begin{abstract}
We analyze theoretically the transport properties 
of a weakly--interacting ultracold Bose gas
enclosed in  two reservoirs connected by a  constriction.  
We assume that  the transport of the superfluid part is
hydrodynamic, and we
describe
the ballistic transport of the normal part using the Landauer-B\"uttiker
formalism. Modeling  the  coupled  evolution  of  the phase,  atom
number, and  temperature mismatches  between the reservoirs, we predict
that Helmholtz (plasma) oscillations, induced  by an initial 
imbalance in atom
numbers, can
be observed  at non--zero temperatures  below $T_c$.
We show that, because of its strong compressibility, the ultracold
Bose gas is
characterized by a fast thermalization compared to the damping time
for plasma oscillations, accompanied by a fast transfer of the normal
component through the constriction. This fast thermalization also
affects the gas above $T_c$, where we present
an explicit comparison to the ideal fermionic case. 
\end{abstract}


\maketitle

Transport without friction is a signature property
of superfluidity, spectacularly illustrated in the fountain effect of
liquid Helium \cite{allen:Nature1938}. Its observation relies on the
use of a superleak, which lets the superfluid through while blocking the 
normal part. Superleaks are familiar elements in the context of 
experiments on liquid helium \cite{donnelly:PhysToday2009}, but their
design in the context of ultracold gases remains an open question. 
Their implementation would allow, for instance, the
implementation of new adiabatic cooling schemes \cite{papoular:PRL2012},
the efficient excitation of second sound \cite{donnelly:PhysToday2009},
and, more generally, an advanced control over transport phenomena.

\begin{figure}
  \centering
  \includegraphics[width=.3\textwidth]
  {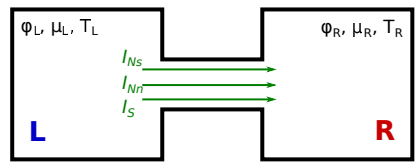}
  \caption{\label{fig:constriction}
    Two reservoirs can exchange particles and heat through a constriction.}
\end{figure}

Recent experiments have initiated the exploration of the
transport properties of ultracold atomic gases
\cite{hazlett:arXiv2013,rancon:arXiv2013,jendrzejewski:arXiv2014,
brantut:Science2012,stadler:Nature2012,brantut:Science2013,lee_SciRep2013}
in geometries comprising two reservoirs
separated by a potential barrier
or by a constriction (see Fig.~\ref{fig:constriction}).
The constriction--based geometry 
is related to those 
investigated in mesoscopic physics \cite{nazarov:QuTransport2009}.
It has already been used to demonstrate
the concept of contact resistance  
\cite{brantut:Science2012},
to investigate superflow \cite{stadler:Nature2012},
and  to observe thermoelectric effects 
\cite{grenier_arxiv2012,brantut:Science2013}
in ultracold Fermi gases.
Superfluids trapped within two connected 
reservoirs are expected to undergo plasma oscillations, which
are analogous to the oscillations of a gas in between two connected
Helmholtz resonators \cite[\textsection 69]{landau:Fluid1987}.
These oscillations have been
extensively studied in the context of liquid helium 
\cite{sato:RepProgPhys2012}.
Similar oscillations have also been observed with
ultracold Bose gases in double--well potentials
\cite{albiez:PRL2005,leblanc:PRL2011}.

In ultracold Fermi gases, the occurrence of BCS--type superfluidity 
occurs at reasonably high temperatures only in the presence of strong
interactions \cite{giorgini:RMP2008}. 
In this case, both the superfluid
and normal parts of the quantum fluid 
are deep in the hydrodynamic regime, which affords a
strong analogy with superfluid 
Helium 
\cite{sidorenkov:Nature2013}. 
However, it also makes it more difficult
to tell the behavior of the superfluid fraction apart from that of the
normal fraction. Hence, 
in the present Letter,
we focus on weakly--interacting bosonic gases, 
where the parameters can be chosen such that superfluid transport
is hydrodynamic whereas normal transport through the channel is ballistic.

We develop a theory describing 
the transport properties
of weakly--interacting uniform Bose gases under these conditions,
reflecting
the different transport regimes for the superfluid and
normal parts.
We use it to show that plasma oscillations are observable 
even at non--zero
temperatures below $T_c$,
and  provide a first description of the damping mechanism due to the
coupling between the superfluid and normal parts.
We also show that the large 
compressibility of the Bose gas leads to surprisingly
fast thermalization compared
to the damping time 
of the transport phenomena. Below $T_c$,
this causes an efficient transport of the normal part at short times;
above $T_c$, it yields a key difference compared to ideal 
fermionic gases.

We describe the ballistic transport of the normal part of the fluid
using the Landauer--B\"uttiker 
formalism for quantum transport \cite[chap.~2]{datta:ElecTransport1995}. 
To our knowledge,
the present work is the first application of this formalism to 
massive bosons.
It had previously
been applied to (massless) phonons to determine the quantum
properties of heat conductance \cite{schwab:Nature2000}.

We assume that the two compartments of Fig.~\ref{fig:constriction} are
box traps with the same volume $V^L=V^R$, each enclosing a 
uniform superfluid. We model the constriction by an isotropic radial
harmonic trap of frequency $\omega_\perp/2\pi$. The hydrodynamic assumption
for superfluid transport through the constriction
\cite[chap.~5]{pitaevskiiStringari:OUP2003} is valid if 
$\hbar\omega_\perp \ll gn$, where $n$ is the mean gas density inside the
constriction, $g=4\pi\hbar^2a/m$ is the interaction constant, $a$ is the
scattering length and $m$ is the atomic mass.

We call $\delta N_s=N_s^R-N_s^L$ and 
$\delta N_n=N_n^R-N_n^L$ 
the difference in superfluid and normal atom numbers between the right
and left compartments of Fig.~\ref{fig:constriction}, and 
$\delta S=S^R-S^L$ the analogous entropy difference. We focus on 
small deviations from the homogeneous situation. In this linear--response
regime, the superfluid current $I_{N_s}$, 
the normal current $I_{N_n}$, 
and the entropy current $I_S$, corresponding to the three differences above, 
are linear functions of the small 
differences in phase $\delta\phi$, chemical potential $\delta\mu$,
and temperature  $\delta T$ between the two reservoirs, which we write
in matrix form as:
\begin{equation} \label{eq:onsagermatrix}
  \begin{pmatrix}
    I_{N_s} \\[6pt]
    I_{N_n} \\[6pt]
    I_{S}/k_B
  \end{pmatrix}
  = 
  \begin{pmatrix}
    2I_J & 0 & 0 \\[6pt]
    0 & L_{11} & L_{12} \\[6pt]
    0 & L_{12} & L_{22}
  \end{pmatrix}
  \begin{pmatrix}
    \hbar\delta\phi \\[6pt]
    \delta\mu  \\[6pt]
    k_B \delta T
  \end{pmatrix}
  \ .
\end{equation}
The first
line in the matrix 
reflects 
the definition of the superfluid current,
$\vec{j}_s=n_s \vec{v}_s$, where 
$n_s$ is the mean superfluid density
in the reservoirs and $\vec{v}_s$ is the superfluid 
velocity. The latter 
satisfies $m\vec{v}_s=\hbar\vec{\nabla}\phi$,
with $m$ being the atomic mass.
For the geometry of Fig.~\ref{fig:constriction},
we find $I_J=n_s A/ml$, 
where $l$ is the constriction length 
and $A=\pi g n_s/m\omega_\perp^2$ is its effective Thomas--Fermi section. 
The two zeroes in the first column reflect the fact that the normal--part
quantities $\delta N_n$ and $\delta S$ do not explicitly depend on the 
superfluid phase difference $\delta\phi$. 
The coefficients $(L_{ij})$ describe the ballistic transport of the normal
part and the entropy.
Assuming that 
$k_B T \gg gn$, the excitations
exchanged by the reservoirs are particles, and an analysis of 
the role of interactions using Hartree--Fock theory reveals
that the ideal--gas expressions for the $L_{ij}$'s are applicable.
This assumption on $T$ 
rules out low--temperature collective phenomena,
such as anomalous phonon transmission \cite{kagan:PRL2003}
or 
Andreev reflection \cite{zapata:PRL2009}.
For uniform Bose gases, this condition
is easy to satisfy
while maintaining the presence of superfluid
($T<T_c$), 
because the ratio $gn/k_BT_c$ is of the order of $0.04$.

We calculate the $L_{ij}$'s using the Landauer--B\"uttiker formalism
\cite[chap.~2]{datta:ElecTransport1995}. We describe the excitations 
in each reservoir using Bose distribution functions $\eta^B$,
whose difference 
$\delta \eta^B=\eta^B_R-\eta^B_L$ satisfies
$\delta \eta^B=
    \partial \eta^B/\partial \mu|_T \, \delta\mu 
   +\partial \eta^B/\partial T|_\mu \, \delta T$.
For temperatures below $T_c$, the $L_{ij}$'s are given by:
\begin{equation} \label{eq:LBcoeffs}
  \begin{split}
    & hL_{11}=-\frac{\pi^2}{6}
    \left(
      \frac{k_B T}{\hbar\omega_\perp}
    \right)^2 
    \ ,
    \\
    & L_{12}=L_{21}=\frac{18}{\pi^2}\zeta(3)L_{11}
    \ ,
    \quad\quad\quad
    L_{22}=\frac{4\pi^2}{5}L_{11}
    \ .
  \end{split}
\end{equation}
The $L_{ij}$'s do not depend on the constriction length 
($\sim 5\,\mathrm{\mu m}$), 
because it is much shorter than
the thermal mean free path inside the reservoirs
($\sim 100\,\mathrm{\mu m}$).
Furthermore, the $L_{ij}$'s 
all share the same dependence on 
$T$ and $\omega_\perp$. This second
property is an important difference with respect to the fermionic case 
\cite{brantut:Science2013}, where the chemical potential is of the
order of the Fermi energy and, hence, enters in the expression for
the transport coefficients.

The coupling between the superfluid and normal parts arises from
the equation of state, which involves the total gas density in each
reservoir, e.g. $n^R=(N_s^R+N_n^R)/V^R$, and from the equation
dictating the evolution of the superfluid velocity,
$\hbar\partial_t\delta\phi=-\delta\mu$
\cite{pitaevskiiStringari:OUP2003}. Combining these equations with
the currents given by Eq.~(\ref{eq:onsagermatrix}), 
we obtain a differential system
describing the evolution of  
$\delta\phi$, $\delta N$, and $\delta T$
\footnote{In Eq.~(\ref{eq:DE_phiNT}), 
we have set to $0$ the top right and lower left matrix coefficients.
These coefficients  
are respectively equal to $\alpha/\kappa$
and $-(\omega_\mathrm{pl} \tau_1)^2 \alpha/(\kappa \ell)$, where
$\alpha=\partial N/\partial T|_\mu$, and 
they have a negligible impact on the dynamics of the system. We
have retained them in our numerical calculations.}%
:
\begin{equation} \label{eq:DE_phiNT}
  \tau_1 \frac{d}{dt}
  \begin{pmatrix}
    \frac{\hbar\delta\phi}{\tau_1} \\[6pt]
    \frac{\delta N}{\kappa_T} \\[6pt]
    k_B \delta T
  \end{pmatrix}
  =
  \begin{pmatrix}
    0 & -1 & 0 
    \\[6pt]
    (\omega_\mathrm{pl} \tau_1)^2 & -1 & +\cal{S} 
    \\[6pt]
    0 
    &
    {\cal{S}}/{\ell} & 
    -\tau_1/\tau_T
  \end{pmatrix}
  \begin{pmatrix}
    \frac{\hbar\delta \phi}{\tau_1} \\[6pt]
    \frac{\delta N}{\kappa_T} \\[6pt]
    k_B \delta T
  \end{pmatrix}
  \ .
\end{equation}
In Eq.~(\ref{eq:DE_phiNT}), $\kappa_T=\partial N/\partial\mu|_T$
is the isothermal compressibility,
$C_N=T\partial S/\partial T|_N$
is the heat capacity, 
$\ell=C_N/\kappa T$ is their ratio, 
and the Seebeck coefficient
$\Scal=-\partial\mu/\partial T|_N-L_{12}/L_{11}$
encodes the thermoelectric properties of the gas.
Equation~(\ref{eq:DE_phiNT}) introduces three timescales:
\begin{equation}
  \label{eq:3timescales}
  \tau_1=\frac{\kappa}{-L_{11}},
  \quad
  \tau_\mathrm{pl}=2\pi\sqrt{\frac{\kappa}{2I_J}},
  \quad
  \tau_T=\frac{C_N/T}{-L_{22}},
\end{equation}
where $\tau_1$ is a damping time associated with normal transport,
the bare plasma period
$\tau_\mathrm{pl}=2\pi/\omega_\mathrm{pl}$ is associated to superfluid
transport, 
and $\tau_T$ is the thermalization time. 
The time $\tau_1$ determines the damping of plasma oscillations and
thermolectric effects.
\begin{figure}
  \centering
  \includegraphics[angle=-90,width=.4\textwidth]
  {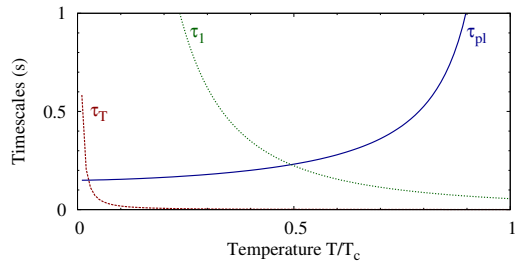}
  \caption{\label{fig:timescales}
    The three timescales $\tau_T$ (dashed red), 
    $\tau_\mathrm{pl}$ (solid blue), 
    $\tau_1$ (dotted green)
    defining the transport of an ultracold ${}^{87}\mathrm{Rb}$ Bose gas.
    Calculated for $N=10^5$ atoms in each reservoir, 
    density $n=10^{19}\,\mathrm{atoms/m^3}$,
    constriction frequency $\omega_\perp/2\pi=15\,\mathrm{Hz}$
    and constriction length $l=5\,\mathrm{\mu m}$.
  }
\end{figure}

\begin{figure*}
  \begin{minipage}{.32\textwidth}
    \centering
    \includegraphics[angle=-90,width=.98\textwidth]
    {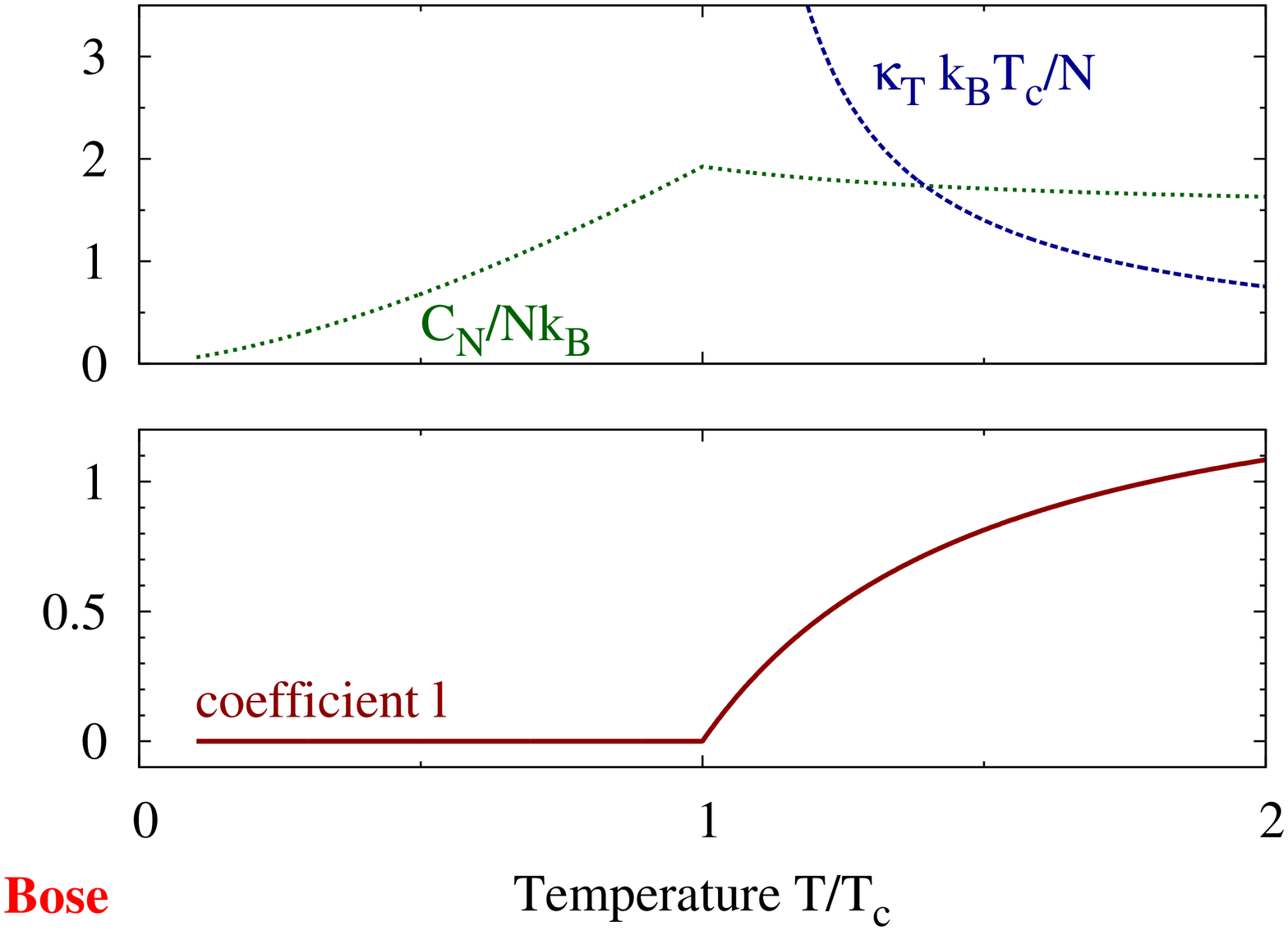}
  \end{minipage}
  \begin{minipage}{.32\textwidth}
    \centering
    \includegraphics[angle=-90,width=.98\textwidth]
    {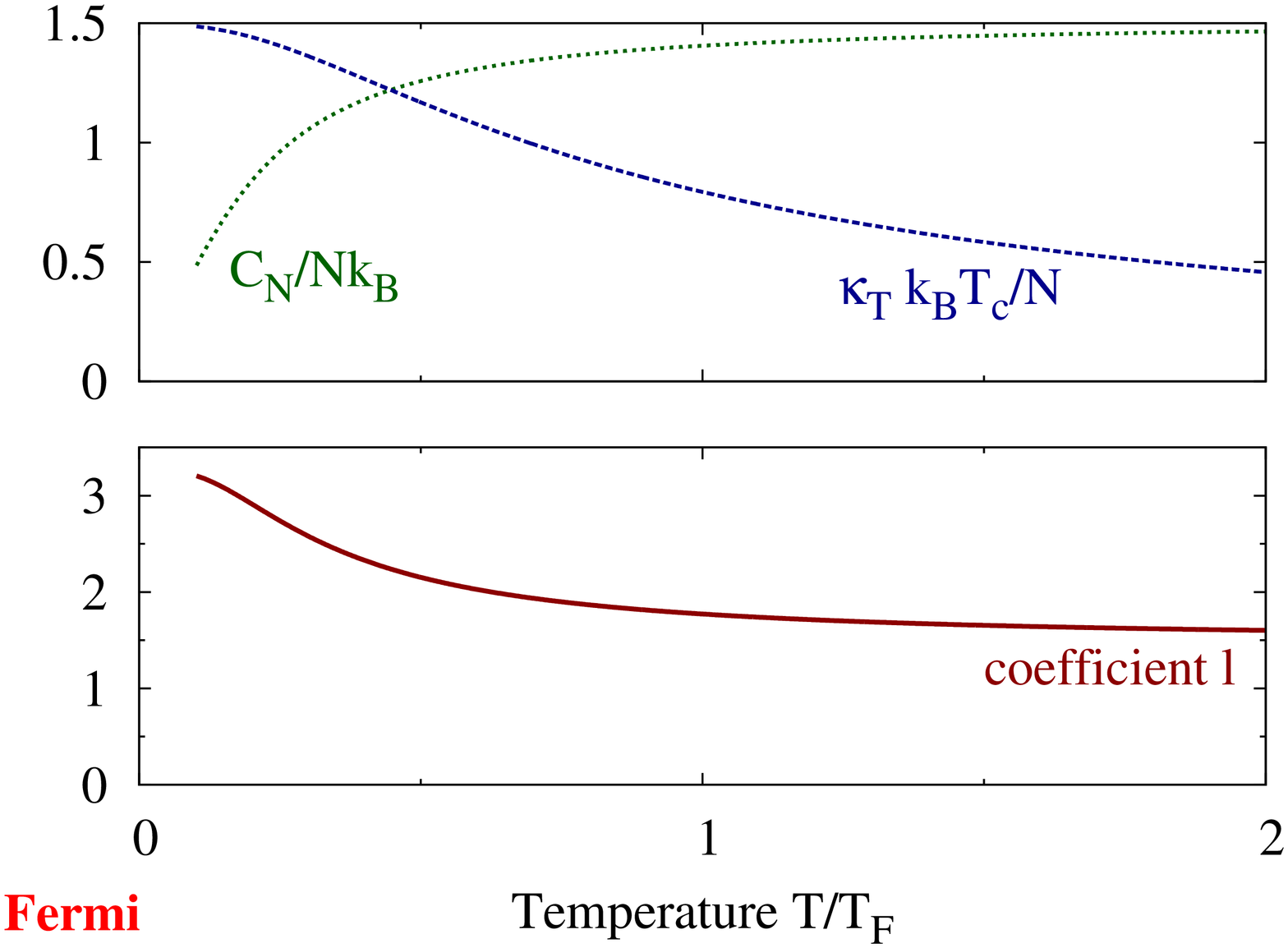}    
  \end{minipage}
  \begin{minipage}{.32\textwidth}
    \centering
    \includegraphics[angle=-90,width=.98\textwidth]
    {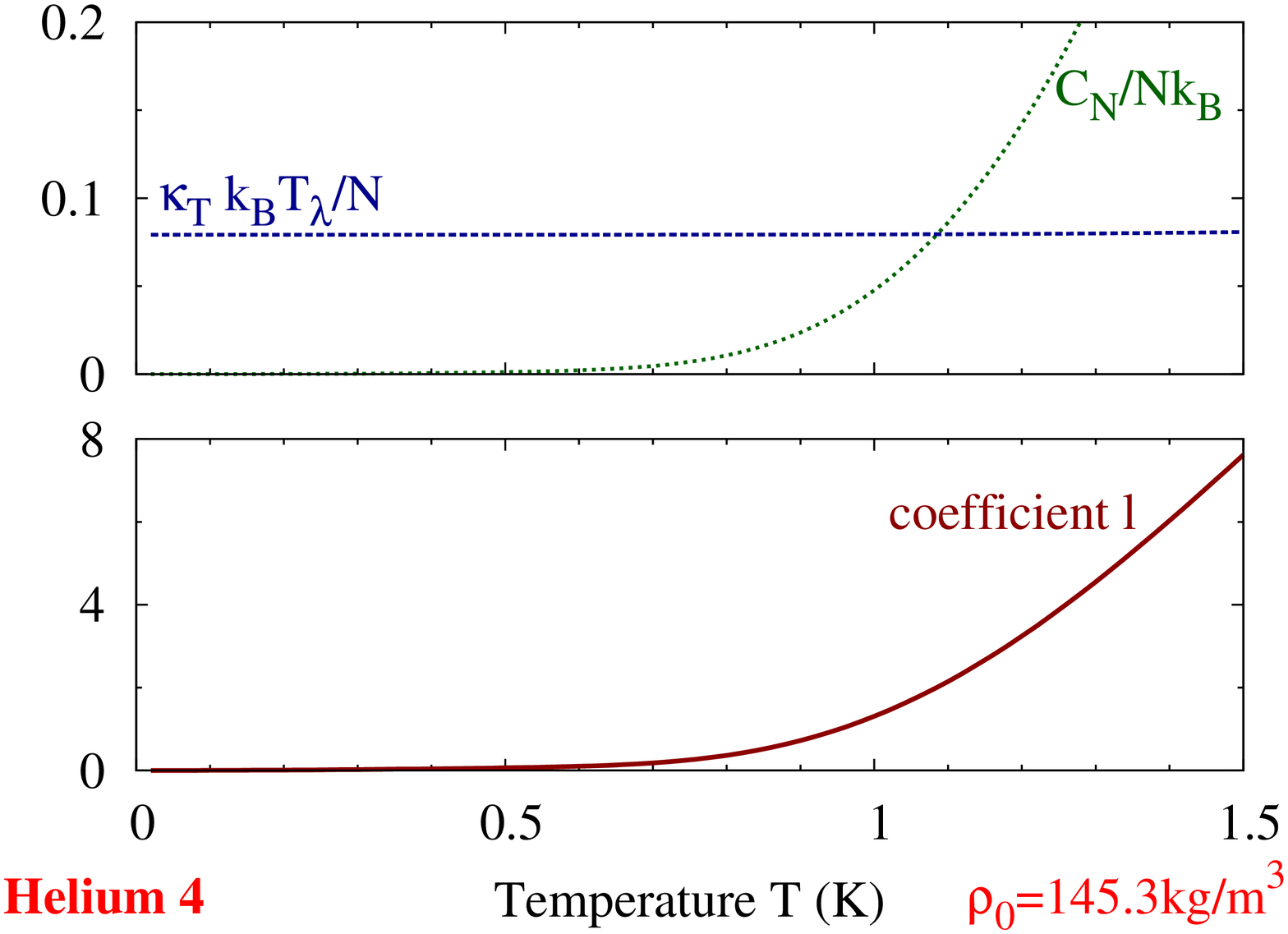}
  \end{minipage}
  \caption{\label{fig:lcal_He4_idB_idF}
  Specific heat $C_N$ (dotted green), 
  compressibility $\kappa_T$ (dashed blue), 
  and their ratio $\ell$ (solid red),
  for ideal Bose (left) and Fermi (center) gases, 
  and for liquid helium 4 
  (right, density $\rho_0=145.3\,\mathrm{kg/m^3}$,
  calculated using the thermodynamic data
  in \cite{arp:IJT2005}).
}
\end{figure*}
Weakly--interacting Bose gases are characterized
by a very large compressibility ($\kappa_T=N/gn$ for $T<T_c$),
whereas $C_N/Nk_B$ remains finite 
(see Fig.~\ref{fig:lcal_He4_idB_idF} left). Hence, the ratio
$\ell$ is very small, of the order of a few $10^{-2}$,
which is a key difference with respect to both ideal Fermi gases
($\ell\sim 1$, Fig.~\ref{fig:lcal_He4_idB_idF} center)
and liquid Helium 4 ($\ell\sim 10$ close to the superfluid transition,
Fig.~\ref{fig:lcal_He4_idB_idF} right). This specific property of Bose 
gases leads to $\tau_T \ll \tau_1$,
i.e. thermalization is much faster than the damping due to
normal transport, as illustrated on Fig.~\ref{fig:timescales}.
Furthermore, for temperatures high enough for Hartree--Fock theory to hold
($k_B T\gtrsim$ a few $gn$), but low enough for the superfluid fraction
$N_s/N=1-(T/T_c)^{3/2}$ to be substantial ($T/T_c\lesssim 0.5$),
the three timescales satisfy $\tau_T \ll \tau_\mathrm{pl} < \tau_1$.
On the other hand, if $T\gtrsim T_c$, the superfluid is absent and
our description reduces to the dynamical system
on $\delta N$ and $\delta T$ introduced in \cite{brantut:Science2013}, 
which corresponds to the lower right $2\times 2$ block of the matrix in
Eq.~(\ref{eq:DE_phiNT}).

\emph{Plasma oscillations.}
We now turn to the analysis of plasma oscillations in
the geometry of Fig.~\ref{fig:constriction}. These oscillations can be
excited by introducing an initial atom number mismatch $\delta N$ between
the two reservoirs. We have predicted their
occurrence at $T=0$ by numerically solving the Gross--Pitaevskii equation,
using a 
Crank--Nicolson scheme \cite{NR3:CUP2007,dalfovo:PRA2000}. We
have investigated a 2D geometry 
inspired by \cite{gaunt:PRL2013}, as well as the cylindrically symmetric 
three--dimensional
geometry corresponding to Fig.~\ref{fig:constriction}. Our results
validate the hydrodynamic approach for superfluid transport and yield
promising orders of magnitude for the plasma oscillation 
frequency $\omega_\mathrm{pl}^{(0)}$, of the order of a few Hz. 
 
\begin{figure*}
  \begin{minipage}{.32\textwidth}
    \centering
    \includegraphics[angle=-90,width=.95\textwidth]
    {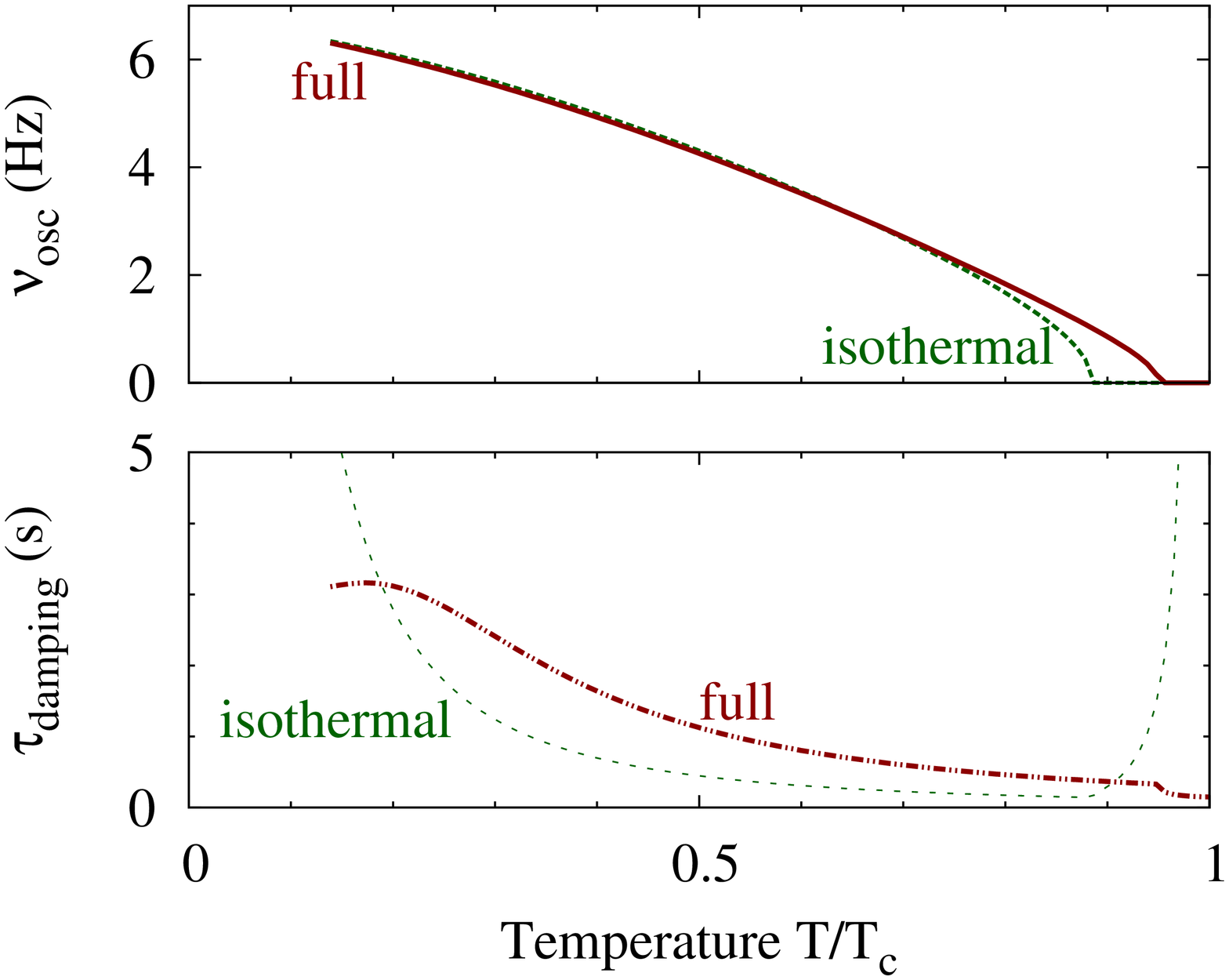}
  \end{minipage}
  \begin{minipage}{.32\textwidth}
    \centering
    \includegraphics[angle=-90,width=.95\textwidth]
    {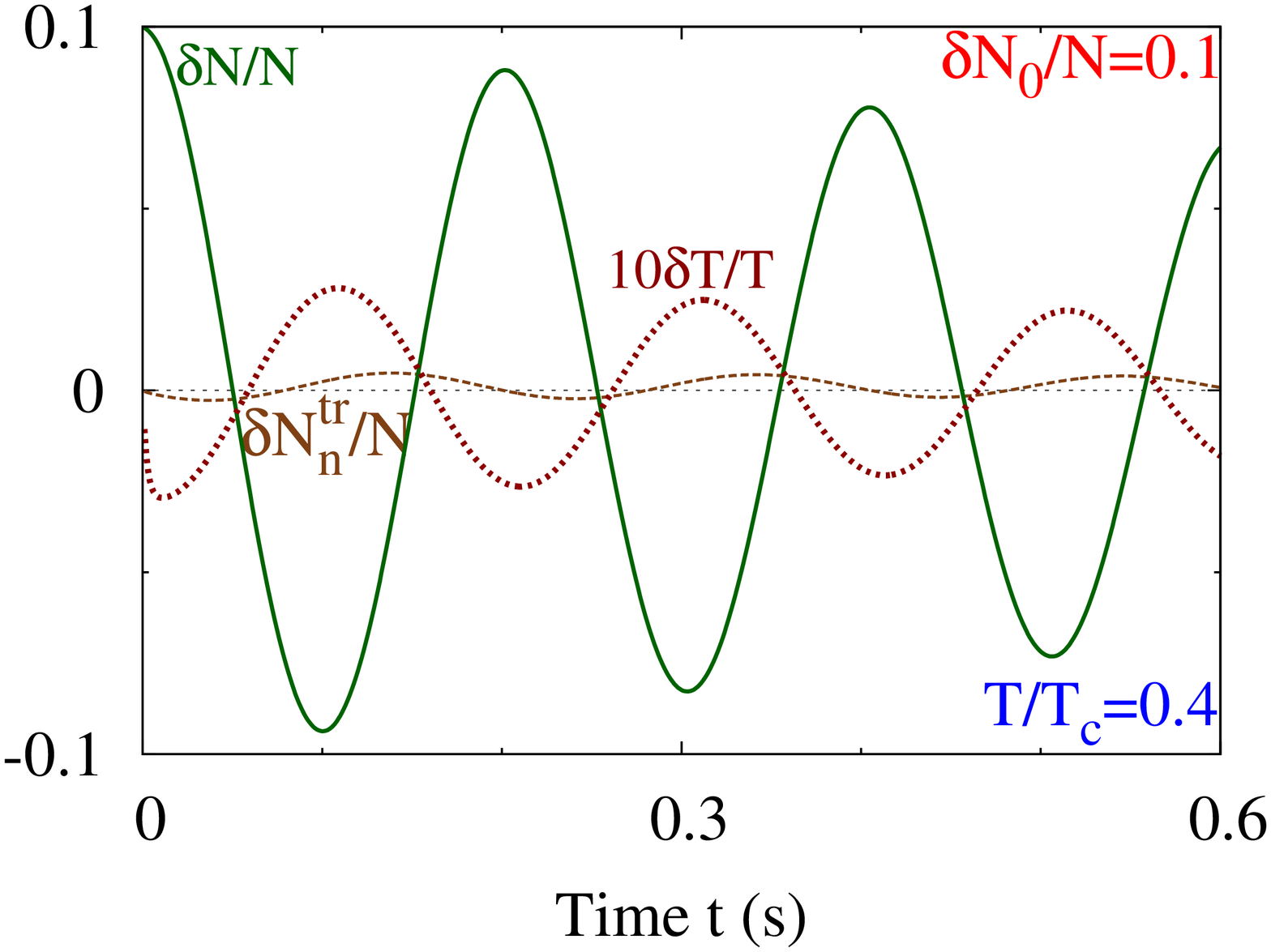} 
  \end{minipage}
  \begin{minipage}{.32\textwidth}
    \centering
    \includegraphics[angle=-90,width=.95\textwidth]
    {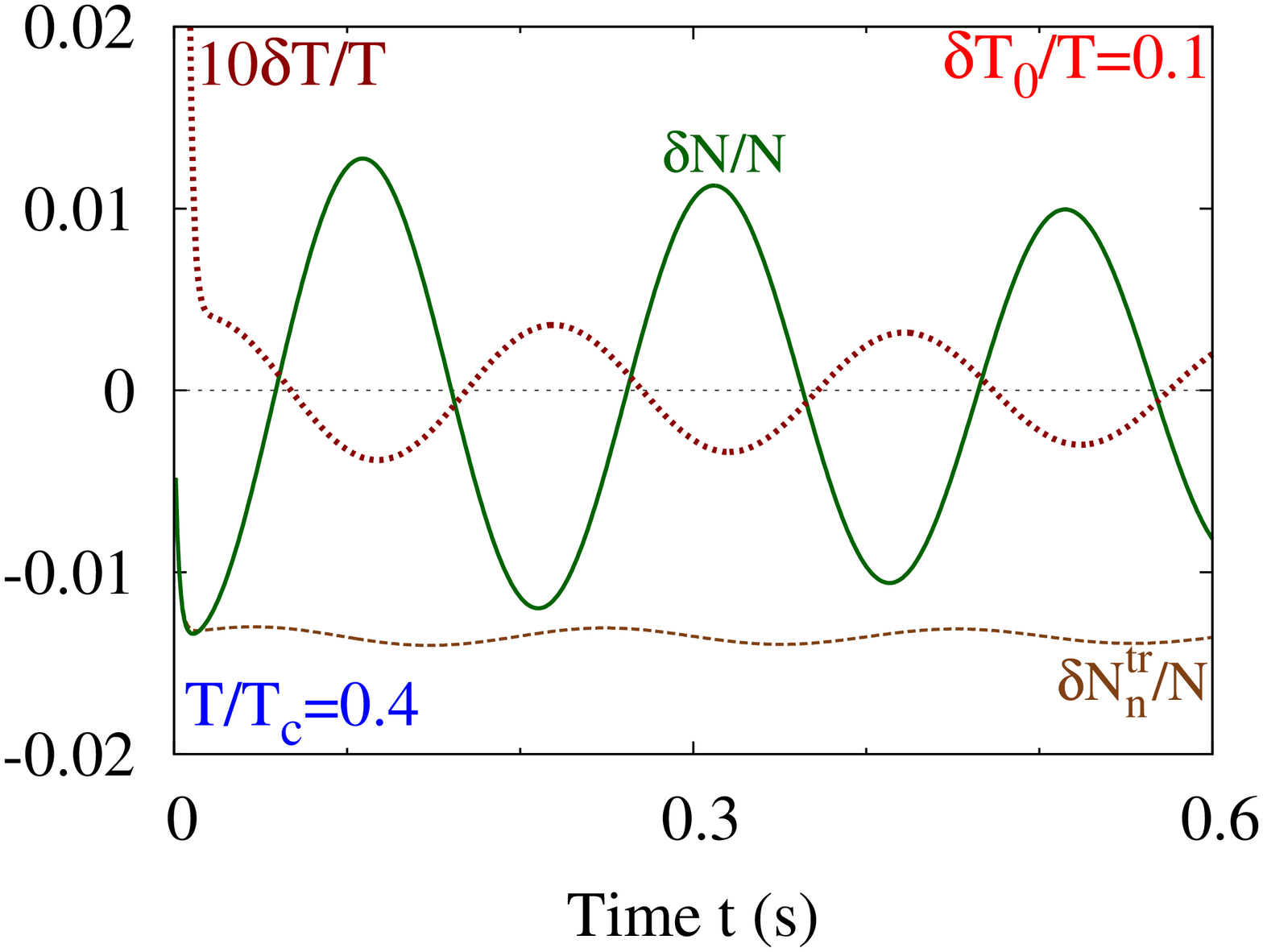}
  \end{minipage}
    \caption{ \label{fig:plasmaosc}
    Superfluid oscillations in an ultracold ${}^{87}\mathrm{Rb}$ Bose gas,
    for the parameters used in Fig.~\ref{fig:timescales}.
    Left:
    frequency $\nu_\mathrm{osc}=\omega_\mathrm{osc}/2\pi$ (top) 
    and damping time $\tau_\mathrm{damp}$ (bottom) for plasma oscillations
    at non--zero temperatures below $T_c$,
    calculated 
    using 
    Eq.~\ref{eq:DE_phiNT} (red ``full'') and 
    its isothermal limit (dashed green ``isothermal'').
    Center: the initial imbalance in atom numbers $\delta N_0/N=0.1$
    causes quasi--isothermal oscillations.
    Right: the initial temperature mismatch $\delta T_0/T=0.1$
    yields fast thermalization accompanied by an efficient transport
    of the thermal part at short times, 
    followed by quasi--isothermal oscillations.
    In both cases, $T/T_c=0.4$, and we plot the time evolution of the
    differences in atom numbers $\delta N/N$
    (solid green), 
    temperature $\delta T/T$ (dotted red, multiplied by 10), and
    transported thermal 
    part atoms $\delta N_n^\mathrm{tr}/N$ (dashed brown).
  }
\end{figure*}

Our model allows us to investigate plasma oscillations
at non--zero temperatures. 
First, our Hartree--Fock description shows that the bare plasma frequency 
$\omega_\mathrm{pl}$ scales with $a/l^{1/2}$, whereas the damping factor
$\omega_\mathrm{pl}\tau_1$ is proportional to $(T_c/T)^2/\sqrt{l}$ and does
not depend on $a$. 
Therefore, the 
observation of
oscillations will be favored by using
smaller constriction lengths, lower temperatures $T/T_c$, and 
larger scattering lengths $a$.
Plasma oscillations occur if the
matrix entering Eq.~(\ref{eq:DE_phiNT}) has two 
complex--conjugate
eigenvalues with negative real parts, 
$(-1/\tau_\mathrm{damp}\pm i\omega_\mathrm{osc})$. 
In this case, the plasma oscillation frequency is $\omega_\mathrm{osc}/2\pi$ 
and the damping time is $\tau_\mathrm{damp}$. 
Figure~\ref{fig:plasmaosc} left 
shows the temperature 
dependence of $\omega_\mathrm{osc}$ and $\tau_\mathrm{damp}$ for a typical
ultracold ${}^{87}\mathrm{Rb}$ Bose gas below $T_c$. 
Oscillations occur for temperatures up to $0.95\,T_c$.
For higher temperatures, the superfluid fraction becomes negligible,
and the damping time coincides with that predicted by the 
normal--part 
model of Ref.~\cite{brantut:Science2013}.

Thermalization being a fast process compared
to the timescales $\tau_\mathrm{pl}$ and $\tau_1$ causes the evolution of
$\delta T$ to approximately decouple from that of
$\delta\phi$ and $\delta N$. Hence, the dynamics of these latter 
two quantities is almost isothermal and is piloted by the upper left
$2\times 2$ block of the matrix entering Eq.~(\ref{eq:DE_phiNT}). 
The maximum amplitude of the 
temperature oscillations can be determined by assuming that the dynamics
of $\delta T$ is driven by that of $\delta N$ and $\delta\phi$:
\begin{equation}
  \label{eq:InitdN0_dTovT}
  \frac{\delta T_\mathrm{max}}{T}=
  \frac{gn}{k_B T} \: 
  \frac{\cal{S}}{{\cal S}^2+L} \:
  \frac{\delta N_0}{N}
  \ .
\end{equation}
The presence of the factor $gn/k_B T$ in Eq.~(\ref{eq:InitdN0_dTovT})
causes $\delta T_\mathrm{max}/T$ to remain
small and confirms the near--isothermal nature of these oscillations.
Figure~\ref{fig:plasmaosc} left shows that the oscillation frequency
and damping time predicted by the isothermal model (green) are in good 
agreement with the full calculation (red). 
Hence,
the main decay mechanism is due to the presence of the normal part.
Thermoelectric effects, neglected in the isothermal model,
are seen to affect mostly the 
damping time, causing it to lengthen. 

The plasma oscillations caused by an initial number mismatch 
$\delta N_0/N=0.1$ are shown on Fig.~\ref{fig:plasmaosc} center, 
for the same parameters
as those used in Fig.~\ref{fig:timescales}. 
This figure also shows the number of normal atoms that
have traveled through the constriction, $\delta N_n^\mathrm{tr}(t)$ %
\footnote{The quantity $\delta N_n^\mathrm{tr}=\int dt I_{N_n}$ differs 
from the
difference in normal atom numbers $\delta N_n$ because in each reservoir
the normal fraction satisfies $N_n/N=(T(t)/T_c(t))^{3/2}$.}%
, to reveal that the plasma oscillations are performed almost exclusively
by the superfluid part.

\emph{Thermalization at temperatures below $T_c$.}
In order to reveal the key role played by fast thermalization in
ultracold Bose gases, we now consider the response of the system 
to an initial temperature mismatch $\delta T_0$.
We consider temperatures $T/T_c\lesssim 0.5$. In this case,
the dynamics of the system at small times of the order of $\tau_T$ 
is driven by the relaxation of temperature towards 
$\delta T=0$. This fast process quickly converts the initial temperature
mismatch $\delta T_0$ into a number imbalance $\delta N_\mathrm{max}$: 
\begin{equation}
  \label{eq:InitdT0_dNovN}
  \frac{\delta N_\mathrm{max}}{N}=
  \frac{15}{4}
  \frac{\zeta(5/2)}{\zeta(3/2)}
  \:
  \frac{{\cal S}}{{\cal S}^2+L}
  \:
  \left(\frac{T}{T_c}\right)^{3/2}
  \:
  \frac{\delta T_0}{T}
  \ .
\end{equation}
The sign of $\delta N_\mathrm{max}$ is dictated by the Seebeck
coefficient $\Scal$, which is negative, just like for fermions
\cite{brantut:Science2013}. 
Furthermore, according to Eq.~(\ref{eq:onsagermatrix}),
temperature variations do not directly couple to the motion of the 
superfluid part.
Hence, this fast relaxation process almost exclusively drives the
transport of normal atoms.
On a longer timescale, the oscillation then proceeds 
quasi--isothermally as before,
with the frequency $\omega_\mathrm{osc}$ and the damping time 
$\tau_\mathrm{damp}$. 
This process is illustrated on  Fig.~\ref{fig:plasmaosc} right,
for 
$\delta T_0/T=0.1$ and 
the parameters used in Fig.~\ref{fig:timescales}.

\emph{Thermalization at temperatures above $T_c$.}
In Bose gases,
the ratio $\ell$ remains small for temperatures 
$T\gtrsim T_c$, where the physics is captured by the ideal--gas
model and a direct comparison with fermions is possible 
(see Fig. \ref{fig:lcal_He4_idB_idF} left and center). 
The gas contains no superfluid part, and the dynamics of $\delta N$ and 
$\delta T$ are described by the lower right $2\times 2$ block of
Eq.~(\ref{eq:DE_phiNT}), which coincides with the model of 
Ref.~\cite{brantut:Science2013}. 
Equation (\ref{eq:3timescales}) shows that the thermalization time 
$\tau_T$ is determined by the specific heat, 
which is of the same order of magnitude for Bose and Fermi gases.
However, the
damping time $\tau_1$ involves the compressibility, which is much larger
for bosons than for fermions. Therefore, damping is much slower in
Bose gases than in Fermi gases.
\begin{figure}
  \centering
  \includegraphics[angle=-90,width=.4\textwidth]
  {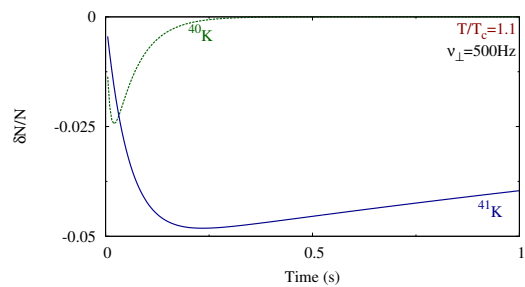}
  \caption{\label{fig:dNovNaboveTc}
  Evolution of $\delta N/N$ following an initial temperature imbalance 
  $\delta T_0/T=0.1$ for bosonic ${}^{41}\mathrm{K}$ 
  (solid blue, $T=1.1 T_c$) 
  and fermionic  ${}^{40}\mathrm{K}$ (dashed green, $T=1.1 T_F$).
  No superfluid is present, and the constriction is more stringent 
  ($\omega_\perp/2\pi=500\,\mathrm{Hz}$) to achieve bosonic 
  decay times of the order of $1\,\mathrm{s}.$
  }
\end{figure}
The variation of $\delta N$ reflects
the two timescales $\tau_T$ and $\tau_T$. In both cases, the 
Seebeck coefficient $\Scal$ is negative, therefore $\delta N$ first 
decreases towards negative values. 
It reaches a minimum for short times 
$t_m \simeq \tau_T\ln\left( \tau_1/\tau_T \right)$,
whose value $\delta N_m=\delta T_0/T \, C_N \Scal/(\Scal^2+\ell)$
does not depend critically on the statistics. However, the
difference between bosons and fermions is apparent during the
long--time relaxation towards $\delta N=0$.  
Figure~\ref{fig:dNovNaboveTc}
compares the cases of bosonic ${}^{41}\mathrm{K}$ at the temperature 
$T/T_c=1.1$ and fermionic ${}^{40}\mathrm{K}$ at the temperature 
$T/T_F=1.1$, with $T_F$ being the Fermi energy.
These two isotopes differ only by the statistics which they obey, 
and the relaxation is 50 times longer for bosons 
($\tau_1^B \sim 1.5\,\mathrm{s}$)
than for fermions  ($\tau_1^F \sim 30\,\mathrm{ms}$).

Strictly speaking, the plasma oscillations we have analyzed
for $T<T_c$ are
not Josephson oscillations. 
These would occur for 
$\mu \ll \hbar\omega_\perp$, which is opposite to
our hydrodynamicity condition for the superfluid flow. 
However, in the 
linear--response limit we envisage, the two equations determining
the superfluid dynamics at $T=0$ 
are formally equivalent to the
Josephson equations \cite[chap.~15]{pitaevskiiStringari:OUP2003}.  
A qualitative difference with true Josephson oscillations will 
emerge in the non--linear regime, where deviations from the law 
$\Delta N_s=2I_J\sin(\Delta\phi)$ should be seen. 
The classical--to--quantum crossover to Josephson oscillations 
can be explored numerically at $T=0$ by varying 
the constriction geometry.

We have shown that, in the case where the transport of the normal part 
through the constriction is  ballistic,
plasma oscillations can be observed at non--zero temperatures below $T_c$,
and that thermalization between the reservoirs is fast compared to the
oscillation period, causing an efficient transport 
of the normal part at short times (see Fig.~\ref{fig:plasmaosc} right).
A possible way to inhibit normal transport, 
and thus to realize a superleak, 
is to add 
a disordered potential inside the constriction, for instance by projecting
a speckle \cite{brantut:Science2013}, in analogy with the fine powders used
in the historical superleaks \cite{allen:Nature1938}. 
The presence of disorder 
should not impede
superfluid flow \cite{lucioni_PRL2011}. 
The anomalously fast thermalization will disappear if
$L_{22}$ is chosen sufficiently small to achieve $\tau_T>\tau_\mathrm{pl}$.

We are grateful to S.~Balibar, I.~Carusotto, G.~Ferrari, and A.~Georges
for fruitful discussions.
This work has been supported by ERC through the QGBE grant.


%

\end{document}